\documentclass[runningheads]{llncs}
\usepackage{graphicx}
\begin{document}
\title{Deep Learning and Random Forest-Based Augmentation of sRNA Expression Profiles}
%
%
\def\and{
  \end{tabular}%
  \hskip 1em \@plus.17fil%
  \begin{tabular}[t]{c}}

\author{Jelena Fiosina\inst{1} \and 
Maksims Fiosins\inst{2,3,4} \and {Stefan Bonn}\inst{2,3}}
\authorrunning{J. Fiosina, M. Fiosins and S. Bonn}
\titlerunning{Deep Learning and Random Forest-Based Augmentation of sRNA}
%
\institute{Clausthal University of Technology, Clausthal-Zellerfeld,
\and
German Center for Neurodegenerative Diseases, T{\"u}bingen,
\and
Institute for Medical Systems Biology, Center for Molecular Neurobiology, University Medical Center Hamburg-Eppendorf, Hamburg,
\and
Genevention GmbH, G{\"o}ttingen, Germany\\
\email{\{jelena.fiosina,maksims.fiosins\}@gmail.com, sbonn@uke.de}}
\maketitle              
\begin{abstract}
The lack of well-structured annotations in a growing amount of RNA expression data complicates data interoperability and reusability. Commonly - used text mining methods extract annotations from existing unstructured data descriptions and often provide inaccurate output that requires manual curation. 
Automatic data-based augmentation (generation of annotations on the base of expression data) can considerably improve the annotation quality and has not been well-studied. We formulate an automatic augmentation of small RNA-seq expression data as a classification problem and investigate deep learning (DL) and random forest (RF) approaches to solve it. We generate tissue and sex annotations from small RNA-seq expression data for tissues and cell lines of {\it homo sapiens}. We validate our approach on 4243 annotated small RNA-seq samples from the Small RNA Expression Atlas (SEA) database. The average prediction accuracy for tissue groups is 98\% (DL), for tissues - 96.5\% (DL), and for sex - 77\% (DL). The "one dataset out" average accuracy for tissue group prediction is 83\% (DL) and 59\% (RF). On average, DL provides better results as compared to RF, and considerably improves classification performance for 'unseen' datasets. 
\end{abstract}
\section{Background}

Qualitative and standartized annotations (tissue, disease, age, sex, cell line, etc.) of  expression data is a key aspect to enable data interoperability and reusability. Data should be findable, accessible, interoperable, and reusable (FAIR), which ultimately facilitate knowledge discovery \cite{Wilkinson16}. Annotations are essential part of semantic data integration systems \cite{Madan18}. In various databases, data annotations are available in different often-unstructured text formats and many times important information on e.g. age, sex, and sometimes even tissue of sample origin is missing (i.e GEO \cite{GEO}). This leads to missing and/or inaccurate annotations, and requires revision and correction by an expert \cite{Hadley18}.
While state-of-the-art expression databases such as the small RNA Expression Atlas (SEA, http://sea.ims.bio) \cite{Rahman17} provide well-structured, ontology-based annotations of publicly available small RNA-seq (sRNA-seq) data, this is achieved by curation of annotations, and missing information is still a problem in many experimental databases. 

A fundamental hypothesis is that augmentation from the source (here, expression counts) data can annotate missing information with high accuracy, allowing for the subsequent analysis of the (meta) data. We suppose that data with similar expression profiles should have similar annotations. Several publications highlight the possibility to use machine learning (ML) approaches to augment expression information, for small RNAs (sRNAs) as well as messenger RNAs (mRNAs). In \cite{Guo17,Hadley18}, the sex in different micro RNA (miRNA) tissues was defined. In \cite{Hadley18}, the authors used the DESeq package and analysis of variance (ANOVA) to detect sex differences in several tissue in miRNAs. In \cite{Ellis18}, age, sex, and tissue were predicted in mRNA sequencing (mRNA-seq) expressions. In \cite{Simon14}, the sex of mRNAs was predicted, and the most important mRNAs were selected. Random Forest (RF) classifier is being widely used 
for classification of expression data, especially in disease diagnostics \cite{Statnikov08}. RF also enables explanation of classification by supplying variable importances.

Deep learning (DL) is making major advances in solving problems that have resisted the best attempts of the artificial intelligence community for many years \cite{Lecun15}. DL is able to deal with big data and is robust even for massive amount of noisy labeled training data \cite{Xiao15}. On the downside, DL requires large amount of training data \cite{Li19}, is prone to overfit on small training sets, and are notoriously hard to biologically interpret (extraction of feature importances) \cite{Webb18}.



In this study we investigate whether the DL-based data augmentation could be superior to classical ML approaches, such as RF. The main hypothesis is that DL classifier trained on sufficiently large data sets would generalize more efficiently to yet unseen 
datasets. Whereas single unseen samples might be easy to learn, datasets usually contain a distinct experimental bias that the model has not learnt a priori.
We apply DL and RF models on human sRNA-seq datasets from SEA, which contains 4243 sRNA-seq samples. Every sample is semantically annotated and analyzed with the same workflow (OASIS \cite{Rahman18}, https://oasis.dzne.de), increasing data interoperability while reducing analysis bias. 


We use this data to predict tissue and sex annotations. DL performs slightly better than RF 
for cross-validataion experiments and significantly overperforms it for
"one dataset out" experiments. 
These results strongly suggest that DL-based expression data augmentation could significantly outperform classical ML approaches, given enough training data.





\section{Methods}
\subsection{Data and Meta-Data Acquisition}

We augment sRNA-seq data with missing annotations. We use SEA sRNA-seq data integration platform that contains 4243 samples and annotations in 350 datasets. The relatively large number of high-quality samples
allows us to use DL for data augmentation purposes. 

We selected 128 {\it homo sapiens} datasets with available annotations for tissue or cell line and sex. We avoided small datasets and samples with rare types of tissues. We used 2806 samples for tissue prediction, including 641 cell line samples with known tissue. For sex classification, we used samples with available sex (only real tissue samples, 1591 samples in 41 datasets). The female and male proportion  was 42\% and 58\%, respectively. We constructed separate classification models for each outcome variable prediction (tissue, sex). 

There are two kinds of expression data available: sRNA expression and the reads not mapped to sRNAs, but mapped to contamination organisms. 
We use both expression profiles, separately and together.
The expression counts from SEA are normalized inside each sample using reads per million (RPM). 

Available tissues are annotated in SEA as specifically as possible. For example, parts of the brain are annotated as "neocortex" or "prefrontal cortex" if this information is available from the experiment. However, using all those tissues in classification leads to a large number of small classes. To avoid this, we joined the available tissues according to used BTO ontology (Table \ref{tab1}). We added also the cell lines to the corresponding groups. We used a hierarchical classification approach: first, we predicted the tissue group and then the single tissue. 

\begin{table}[h]
\caption{Tissue and cell line grouping according to ontologies.}\label{tab1}
\begin{tabular}{|p{0.2\textwidth}|p{0.8\textwidth}|}
\hline
{\bf Tissue group}&{\bf Containing tissues}\\
\hline
blood\_group & blood, blood plasma, blood serum, peripheral blood, umbilical cord blood, serum, buffy coat, immortal human B cell, liver, lymphoblastoid cell\\
brain\_group & brain, cingulate gyrus, motor cortex, prefrontal cortex, neocortex\\
epithelium\_group & skin, dermis, epidermis, breast, oral mucosa, larynx\\
gland\_group & prostate gland, testis, kidney, bladder, uterine endometrium, tonsil, lymph node\\
intestine\_group & intestine, colon, ileal mucosa\\
\hline
\end{tabular}
\end{table}

\subsection{Data scaling and filtering}


\subsubsection{Data Scaling (DL only):} We scaled counts of each sRNA independently. We compared two alternative scalers. A MinMax scaler scales the data in the range (0,1). A standard scaler standardizes features by removing the mean and scaling to unit variance. The MinMax scaler showed better results.

\subsubsection{sRNA filtering (for both RF and DL):} The number of features (sRNAs) was considerably greater than the number of available observations (samples). The initial number of factors was approximately 35000, while the number of available samples was 2200 (for tissue prediction, even fewer for sex). In addition, approximately 5600 contamination counts were available for each sample. 

Most of the counts were equal to zero. 
The preliminary experiments showed that the maximal accuracy was obtained by excluding variables (sRNAs and contaminants) containing more than 30\% of zeroes. After this, the number of sRNAs and contaminants was approximately 2500 and 2000 respectively.

\subsubsection{Sample filtering (for both RF and DL):}\label{sampleFilterSec} Some tissues we could not group (i.e. milk, urine, heart, etc), especially if they were presented in only one dataset. This made "one dataset out" classification (s. Section \ref{validationSec}, Validation) impossible, and we did not predict tissue in such datasets. We also excluded some tissues and cell lines that were presented in one dataset containing less than 9 samples. The cell lines located in the t-distributed stochastic neighbor embedding (t-SNE) plot in other region as the corresponding tissue, were also excluded. The reason for this is that
such cell lines are not similar to original tissue and should be predicted separately. 

After this exclusion, 105 datasets are left, containing 2215 samples. The proportions of cell lines and tissue samples are 23\% and 77\%, respectively. Fig.~\ref{fig1} illustrates the t-SNE plot for the tissue groups.

\begin{figure}[h]
\centering
\includegraphics[width=0.9\textwidth]{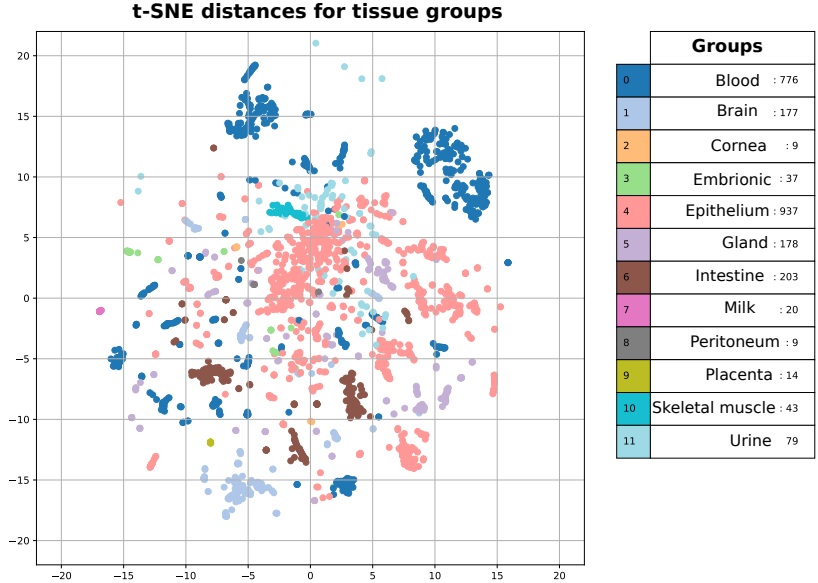}
\vspace{-0.7cm}
\caption{t-SNE Plot for available tissue types.} 
\label{fig1}
\end{figure}

\subsection{Models}
\subsubsection{DL model:}

We used a fully connected neuronal network (NN) architecture. It has one input layer with number of inputs equal to the number of
variables after the initial filtering; we tested the NN with different hyper-parameters (such as layer sizes, number of layers, and drop-out rates). Finally, we used a NN with three hidden layers containing 1000, 250, and 250 neurons, with the drop-out rates 0.5, 0.4, and 0.4. The number of neurons in the output layer was equal to the number of predicted classes. 

We examined different optimizers: 'rmsprop', 'adam', 'sgd', 'adadelta'. We used the rectifier linear unit (ReLU) activation function for our initial and hidden layers. We chose the "softmax" activation for multi-class classification. 

We trained the NN for 50 epochs with batch size 30.
 

\subsubsection{RF model:}

On both stages of the RF, the following parameters were used: mtry equal to the square root of the number of features, and down-sampling to balance the imbalanced classes (especially for tissue prediction). On the first stage (pre-classification), the RF was based on all filtered columns, and the number of trees was 100. We ordered the features according to their importance (Gini index decrease). We used the top-1000 selected features for the second stage classification with an increased number of trees (here, 500). We used RF models for obtaining variable importances

\subsubsection{Validation:}
\label{validationSec}

We implemented two types of cross-validation to check the accuracy of data augmentation in two different conditions.
First, we used 5 fold CV as one readout, where we reported the average performance.
Second, we trained a model using CV and classified a test dataset, which was not seen by the model during training. More specifically, this data came from a different experiment, which contained a different bias, but had a tissue that the model was trained on. This case is more relevant to the real situation, because for automatic augmentation, one should augment the new dataset, taking other datasets as training data. In the case of tissue prediction, such a validation technique was not available for each dataset, because some tissues are available in one dataset only. Throughout this manuscript we will refer to the 5 fold CV as 'cross validation' and the validation on unseen datasets as 'one dataset out'.


\subsubsection{Quality metrics:}
The main metric of the classification quality was accuracy.
Apart from the accuracy, we used various other metrics 
such as: confusion matrix, precision, recall, F1 score in macro and micro versions, and Cohen’s kappa, which normalizes the accuracy by the imbalance of the classes in the data. Those metrics we used internally to tune the classification models.

\subsubsection{Software libraries:}
All the scripts for DL classification are developed in R based on the "keras" library. The RF models are also implemented in R, using the "randomForest" library. We used the Python 3.5 "sklearn.manifold" t-SNE library to build the t-SNE plots.

\section{Results}

The main hypothesis of this study is that DL-based expression data augmentation approaches might outperform classical approaches. We therefore compared DL to RF classification to predict the target tissue and patient sex of human sRNA-seq data.
A second aim was to analyse variable importance to check their biological relevance.

\subsection{Robust sRNA-seq tissue prediction}
\subsubsection{Tissue group prediction:}

\paragraph{CV experiments}
We experimented with 9 and 15 minimal number of samples per tissue class.  Fig. \ref{fig2} (left) shows that RF is less accurate, especially for a class with a minimum of 9 samples: DL: 97\%, RF: 85\%. For the classes with a minimum of 15 samples, the accuracy was better: DL: 98\%, RF: 92\%. The DL model gave better results, in both cases, because it did not suffer from imbalanced classes, however we used an internal class balancing mechanism for the RF model.

\begin{figure}[h]
\includegraphics[width=\textwidth]{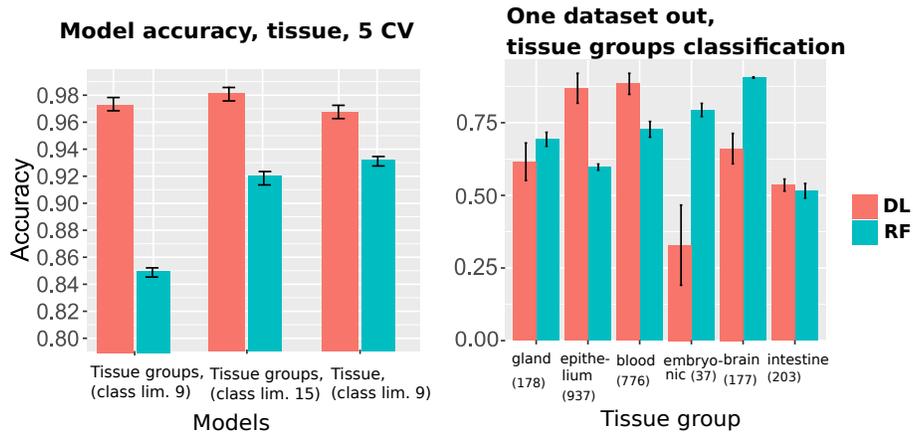}
\caption{CV tissue pred. accuracy (left); "One dataset out" tissue groups pred. accuracy (right)}
\label{fig2}
\end{figure}

\paragraph{"One dataset out" experiments} After initial filtering only 6 aggregated tissues were left. The reason was that 
some tissues were only in one dataset and some tissues were presented in datasets containing less than 9 samples (see Sec. \ref{sampleFilterSec}, Sample Filtering).
In Fig. \ref{fig2} (right), we present the accuracy of tissue group prediction. Notice that the datasets with the same tissue may differ from dataset to dataset because of different factor influences (e.g., library preparation methods, biological conditions of samples: cell types, diseases). This is a reason for the significantly lower model accuracy in this case. 
For the intestine group, which we could not detect very well, the accuracy was around 50\%. We could predict most of the datasets with accuracy of 80-100\%. 
The average accuracy is 83\% (DL) and 59\% (RF). 



\subsubsection{Tissue prediction:}\label{tissuePredSec}
\paragraph{CV experiments}
We avoided combining any tissue or cell line. Instead, we used all the tissue and cell line classes that had more than nine samples. The DL und RF models had standard parameters, as described above. The average accuracy (Fig. \ref{fig2}, left) was DL: 96.5\%, RF: 93\%. The classes were not as imbalanced without tissue aggregation, and thus we got similar results with both models.

\paragraph{"One dataset out" experiments} Knowing the tissue group from the previous experiments, we specified its tissue class. The dataset exclusion criteria were the same as in previous experiments with tissue groups. The resulting histogram for each dataset as a test set are presented in Fig. \ref{fig3} (left). The tissue in the most datasets is predicted within the accuracy interval (0.8,1); nevertheless, tissue in some datasets is predicted with accuracy (0,0.2). In Fig. \ref{fig3} (right) and then in Fig. \ref{fig4}, we can see the tissues and cell lines that could not be predicted well (brain, breast, colon, skin, etc.).
Bad "brain" tissue prediction was caused by its identification as sub/tissues: prefrontal cortex and neocortex.
It could be true, because the sub-tissue had no annotation in the given dataset.

\begin{figure}[h]
    \begin{minipage}{0.4\textwidth}
\includegraphics[width=\textwidth]{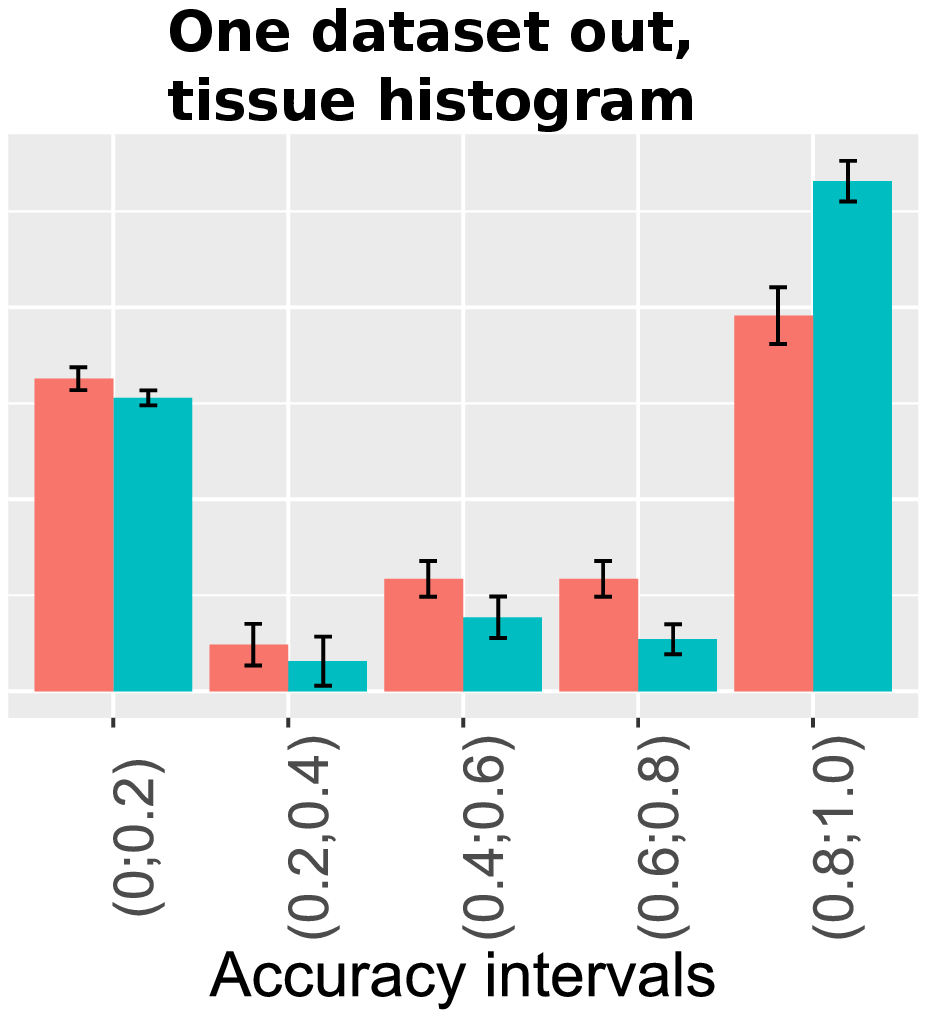}
    \end{minipage}\hfill
    \begin{minipage}{0.6\textwidth}
\includegraphics[width=\textwidth]{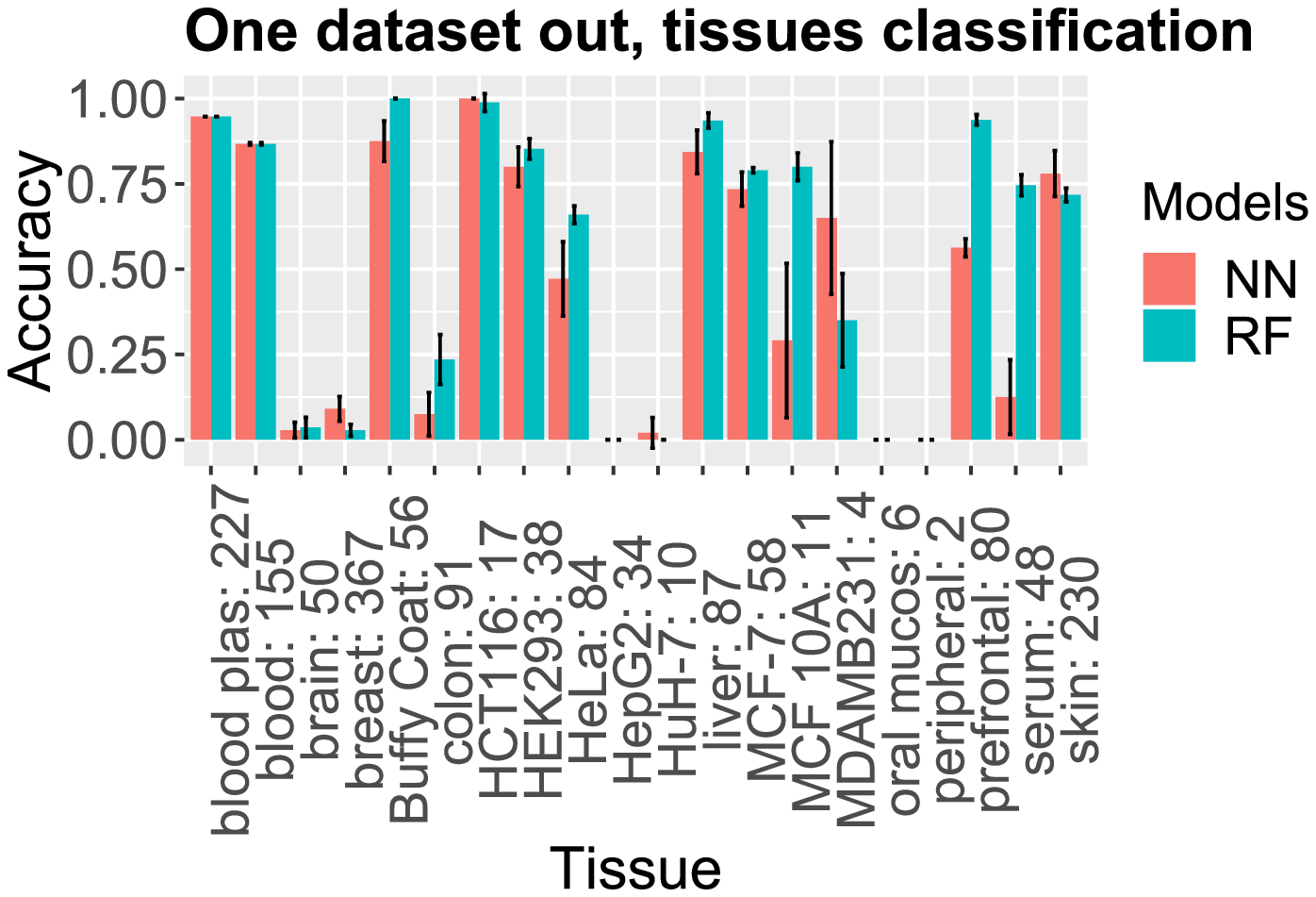}
    \end{minipage}\hfill
\caption{"One dataset out" tissue pred. accuracy histogram; "One dataset out" tissues pred. accuracy by classes.} \label{fig3}
\end{figure}

\begin{figure}[h]
\includegraphics[width=\textwidth]{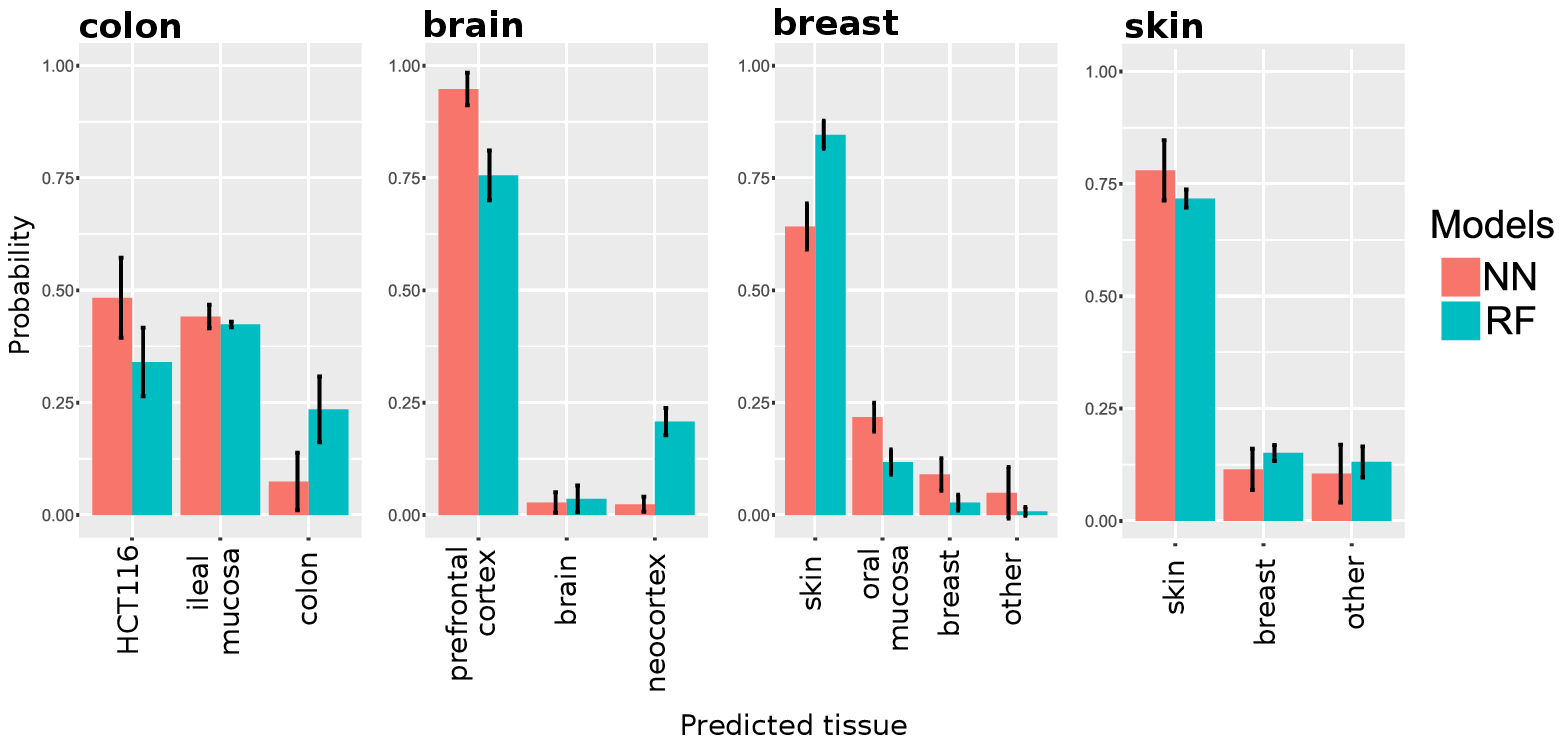}
\caption{"One dataset out" of tissue groups each dataset pred. accuracy.} \label{fig4}
\end{figure}

"Breast" and "skin" tissues are very similar, and both were not identified correctly in many datasets. "Colon" tissue was identified as the "HCT116" cell line, i.e., colon cancer and as "ileal mucosa", which is very near to the colon. 

We conclude that for tissue group prediction DL overperforms RF, especially in "one dataset out" case. For tissue prediction the 
difference was smaller, but DL was still better.

\subsection{Robust sRNA-seq sex prediction}\label{sexPredSec}

The DL and RF models had standard parameters described above. To improve the model accuracy apart from sRNA-seq expression counts, we extend the models with contamination expression counts (Fig. \ref{fig5}). 

\begin{figure}[h]
\includegraphics[width=\textwidth]{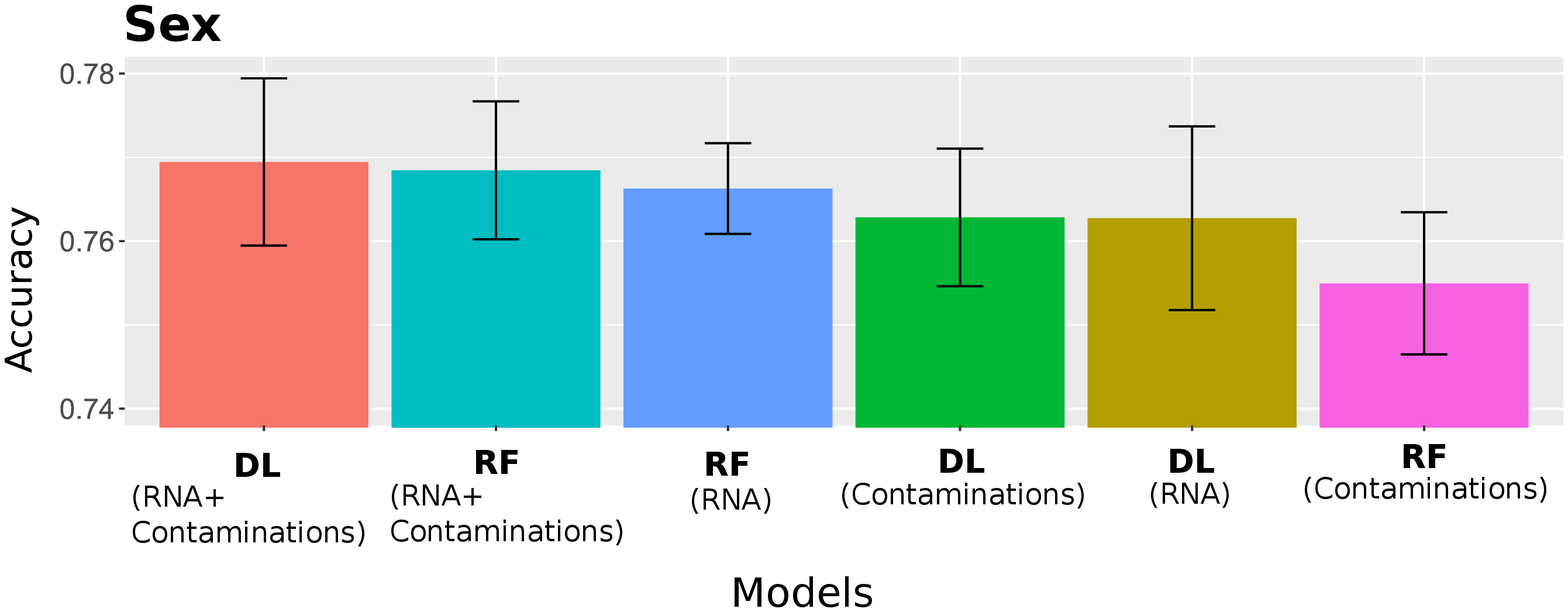}
\caption{CV sex prediction accuracy with different models.} \label{fig5}
\end{figure}

The best models were DL and RF based on both sRNAs and contaminations, with accuracies of 77\% and 76.9\%. The other three models RF(RNA), DL(contaminations), and DL(RNA) gave an accuracy of approximately 76.2\%. It was unexpected that the model based on contaminations only could predict the sex with an accuracy of approximately 76\% for both DL and RF. So for sex prediction DL slightly overperforms RF.

\section{Enrichment tests}

Given the good prediction accuracy we next investigated whether the ML models learn tissue- and/or sex-specific sRNAs. The hypothesis is that to govern accurate prediction the model has to put more emphasis on sRNAs that contain biologically relevant information, in a given context, while ignoring non-relevant information. For the tissue prediction use case, this would imply that a good classifier would put heavy emphasis on sRNAs that are tissue-specific whereas it would put little weight on house-keeping sRNA expression, which is largely invariant over tissues. The same should be true for the sex prediction.

We used the miRNA enrichment analysis and annotation (miEAA) tool \cite{Backes16} and run over-representation analysis with default settings, no reference miRNA set and checking Organs, Diseases and Age/Gender dependent miRNAs. The tool was developed for miRNAs, so we excluded other types of sRNAs from the analysis. We performed the enrichment test on prediction of tissue groups and sex. We took the top-200 miRNAs from the RF classifier (Sec. \ref{tissuePredSec}-\ref{sexPredSec}).

\subsection{Tissue-specific sRNA enrichment}

First, we investigated the enrichment of biological categories for miRNAs that are important for tissue classification. In Fig. \ref{fig6} we see the enrichment of stem cells responsible for tissue-specific tissue formation, and of the cytoskeleton. Blood (including lymphocytes) and adipose tissue show some tissue-specific categories. 
However, the full set of top miRNAs would probably not provide a clear enrichment, because the miRNA subsets used by the classifier to detect the particular tissue groups are mixed.

\begin{figure}[h]
\includegraphics[width=\textwidth]{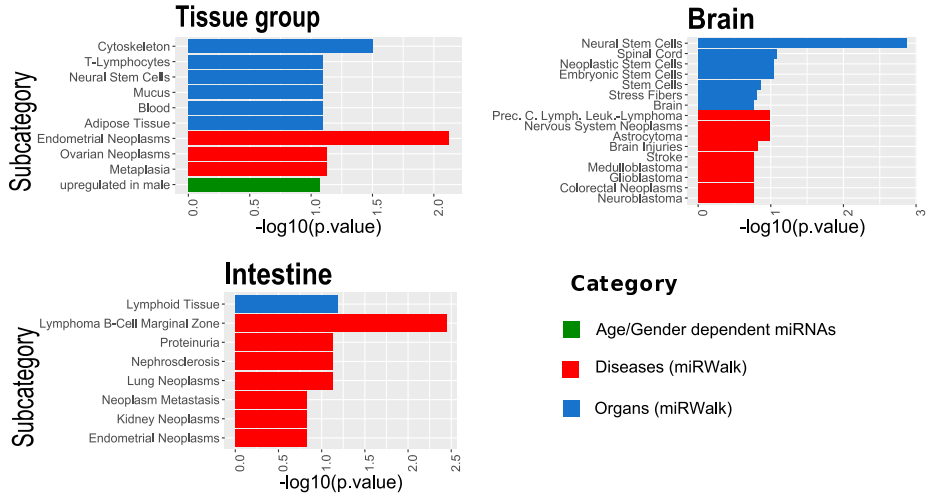}
\caption{Enrichment results for tissue-specific categories.} \label{fig6}
\end{figure}

Next, we clustered sRNAs by their expression.
We could see specific clusters of tissue groups with highly expressed miRNAs. Next, we analyzed each specific cluster separately. 


\subsubsection{Brain:} The cluster contains 43 miRNAs. 10 of them (miR-124, miR-128, miR-129, miR-137, miR-138, miR-153, miR-323, miR-708, miR-99, and miR-9) are reported as brain-specific in \cite{Guo14}. The enrichment test (Fig. \ref{fig6}) shows that most of the enriched categories are brain-specific or nervous system-specific. Therefore, this cluster is well-suited for detection of the brain group.

\subsubsection{Intestine:} This cluster contains 22 miRNAs. 3 of them (miR-10a, miR-196, and miR-200a) are reported as kidney-specific, and one(miR-196) as liver-specific in \cite{Guo14}. In addition, four of them (miR-192, miR-194 and miR-215) are reported as kidney-specific in \cite{Sun04}. Moreover, miR-31 is reported as brain-specific in \cite{Guo14}.The enrichment test (Fig. \ref{fig6}) shows, from organ/tissue category, that the lymphoid tissue is enriched, and may be associated with intestine. 
Therefore, this cluster in general suits for detection of the intestine group.


\subsubsection{Blood:} This cluster contains only six miRNAs. Four of them (miR-129, miR-9, miR-323 and miR-708) are reported as brain-specific in \cite{Guo14}. However, miR-129 is a candidate biomarker for heart failure, and thus is heart/blood specific. The set of six miRNAs is too small for the enrichment test. The classifier uses this cluster more for brain detection than for blood detection, as some sRNAs are highly expressed both in blood and in brain.

The results indicate that the ML learns relevant tissue-specific sRNAs, especially for the brain and intestine clusters.

\subsection{Sex specific sRNA enrichment}

We investigated the enrichment of biological categories coming from sex classification. 
Fig. \ref{fig8} (left) illustrates enrichment of sex-specific terms (upregulated in male, sex-dependent). A broad range of tissues (liver, kidney, adipose tissue, serum, skeletal muscle, bones, breast, and ovary) is enriched. This may show a sex specificity of miRNA expression in many organs. 

\begin{figure}[h]
\centering
\includegraphics[width=0.8\textwidth]{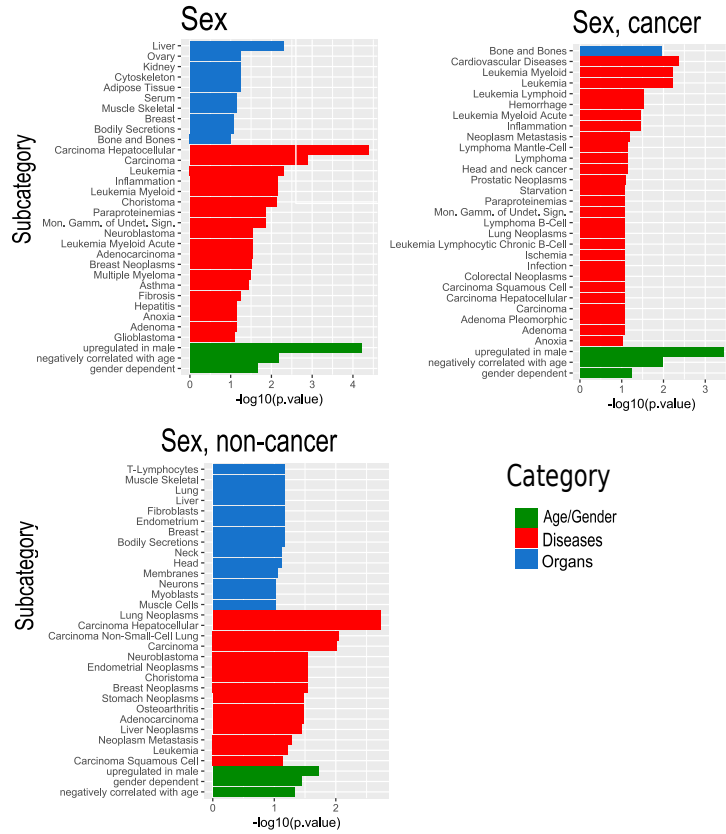}
\caption{Enrichment results.}
\label{fig8}
\end{figure}

The list of enriched diseases mostly contains cancer. Considering that cancer is more frequent in males (approx. 1.5 times), we checked whether the classifier had learned disease instead of learning sex. We divided all available samples into cancer-specific and non-cancer specific, and classified sex separately (Fig. \ref{fig8}).

For cancer-related samples, a similar list of diseases is enriched. Therefore, this classifier may be disease-based. However, for non-cancer samples, tissues and a shorter list of diseases are enriched. Therefore, this classifier is really based on organ sex specificity. This means that the classifier based on all samples causes mixed sex classification: sometimes based on disease, sometimes based on organs. 

\section{Conclusion and future work}

We compared the performance of DL with classical (RF) approach for  prediction of tissue and sex based of human sRNA-seq expression data.
The obtained results show that DL based augmentation overperforms RF,
especially in the 'one dataset out' validation. DL acts as a "black box" model, while  RF allows to explain variable importance. 
As our future work, we are going to predict age in the same manner as sex, improve our models by stacking a combination of various models, 
and apply our models for other types of expression data. We are also planning to conduct more accurate variable and sample filterning, as well as more deep result interpretation, including enrichment of non-miRNA sRNAs and contaminants. 

\subsubsection{Acknowledgements:} The  research was supported by the  German  Federal  Ministry  of  Education  and Research (BMBF), project Integrative Data Semantics for Neurodegenerative research (031L0029);  by German Research Foundation (DFG), project Quantitative Synaptology (SFB 1286 Z2) and by Volkswagen Foundation.

\bibliography{FiosinaFiosins_ISBRA}{}

\begin{thebibliography}{10}
\providecommand{\url}[1]{\texttt{#1}}
\providecommand{\urlprefix}{URL }
\providecommand{\doi}[1]{https://doi.org/#1}

\bibitem{Backes16}
Backes, C., Khaleeq, Q.T., et~al.: mieaa: microrna enrichment analysis and
  annotation. Nucleic acids research  \textbf{44}(W1) (2016)

\bibitem{Ellis18}
Ellis, S., et~al.: Improving the value of public rna-seq expression data by
  phenotype prediction. Nucleic Acids Res.  \textbf{46}(9) (2018)

\bibitem{GEO}
Gene expression omnibus, \url{https://www.ncbi.nlm.nih.gov/geo/}

\bibitem{Guo17}
Guo, L., et~al.: mirna and mrna expression analysis reveals potential
  sex-biased mirna expression. Scientific Reports  \textbf{7} (2017)

\bibitem{Guo14}
Guo, Z., Maki, M., et~al.: Genome-wide survey of tissue-specific microrna and
  transcription factor regulatory networks in 12 tissues. Scientific Reports
  \textbf{4} (2014)

\bibitem{Hadley18}
Hadley, D., Pan, J., et~al.: Precision annotation of digital samples in
  ncbi’s gene expression omnibus. Sci. Data  \textbf{4},  170125 (2017)

\bibitem{Lecun15}
LeCun, Y., Bengio, Y., Hinton, G.: Deep learning. Nature  \textbf{521} (2015)

\bibitem{Li19}
Li, Y., et~al.: Deep learning in bioinformatics: introduction, application, and
  perspective in big data era. bioRxiv  (2019)

\bibitem{Madan18}
Madan, S., Fiosins, M., et~al.: A semantic data integration methodology for
  translational neurodegenerative disease research. figshare  (2018)

\bibitem{Rahman17}
Rahman, R.U., Sattar, A., Fiosins, M., et~al.: Sea: The small rna expression
  atlas. bioRxiv  (2017),
  \url{https://www.biorxiv.org/content/early/2017/08/04/133199}

\bibitem{Rahman18}
Rahman, R.U., et~al.: Oasis 2: improved online analysis of small rna-seq data.
  BMC Bioinformatics  \textbf{19}(54) (2018)

\bibitem{Simon14}
Simon, L., et~al.: Human platelet microrna-mrna networks associated with age
  and gender revealed by integrated plateletomics. Blood  \textbf{123}(e37-e45)
  (2014)

\bibitem{Statnikov08}
Statnikov, A., Wang, L., Aliferis, C.F.: A comprehensive comparison of random
  forests and support vector machines for microarray-based cancer
  classification. BMC Bioinformatics  \textbf{9} (2008)

\bibitem{Sun04}
Sun, Y., Koo, S., et~al.: Development of a micro-array to detect human and
  mouse micrornas and characterization of expression in human organs. Nucleic
  Acids Res  \textbf{32}(22) (2004)

\bibitem{Webb18}
Webb, S.: Deep learning for biology. Nature  \textbf{554},  555--557 (2018)

\bibitem{Wilkinson16}
Wilkinson, M.D., et~al.: The fair guiding principles for scientific data
  management and stewardship. Sci. Data  \textbf{3},  160018 (2016)

\bibitem{Xiao15}
Xiao, T., et~al.: Learning from massive noisy labeled data for image
  classification. In: 2015 IEEE Conf. on Comp. Vision and Pattern Recognition
  (CVPR). pp. 2691--2699 (2015)

\end{thebibliography}
\bibliographystyle{splncs04}
\end{document}